Benhai Wang, Haobin Han, Lijun Yu, Yueyue Wang[*], Chaoqing Dai[*]

# Generation and dynamics of soliton and soliton molecules from a VSe₂/GO-based fiber laser

**\*Corresponding author: Chaoqing Dai, Yueyue Wang,** College of Optical, Mechanical and Electrical Engineering, Zhejiang A & F University, Hangzhou, 311300, China. e-mail: dcq424@126.com; yyshiyan@126.com.

**Benhai Wang, Haobin Han, Lijun Yu,** College of Optical, Mechanical and Electrical Engineering, Zhejiang A & F University, Hangzhou, 311300, China.

**Abstract:** Recently, in addition to exploring the application of new saturable absorber devices in fiber lasers, soliton dynamics has also become a focus of current research. In this article, we report an ultrashort pulse fiber laser based on VSe₂/GO nanocomposite and verify the formation process of soliton and soliton molecules by the numerical simulation. The prepared VSe₂/GO-based device shows excellent saturable absorption characteristics with a modulation depth of 14.3% and a saturation absorption intensity of $0.93MW/cm^2$. The conventional soliton is obtained with pulse width of 573fs, which is currently the narrowest pulse width based on VSe₂-related material, and has a signal-to-noise ratio of 60.4 dB. In addition, the soliton molecules are realized based on the VSe₂/GO for the first time and have a pulse interval of ~2.2ps. We study the soliton dynamics through numerical simulation and reveal that before the formation of the soliton, it undergoes multiple nonlinear stages, such as soliton mode-locking, soliton splitting, and soliton oscillation. Furthermore, the results of numerical simulation are agreed well with the experimental data. These results indicate that the VSe₂/GO might be another promising saturable absorber material for ultrafast photonics, and deepen the understanding of soliton dynamics in ultrafast fiber lasers.

## 1 Introduction

Recently, ultrashort pulse fiber lasers had essential applications in the fields of micro-nano manufacture, optical modulators, three-dimensional optical tweezers, and optical communication [1-3]. Passively mode-locked (PML) technology based on two-dimensional(2D) saturable absorbers (SA) is a necessary technical means for generating ultrashort pulses [4-7]. Due to the advantages of compact structure, stable and alignment-free operation, PML fiber lasers have been widely studied and become an ideal platform for the study of new pulse dynamics. In this platform, in addition to studying conventional soliton pulses, new nonlinear phenomena such as soliton molecules, soliton explosions, and soliton distillation can be studied [8]. Graphene has been successfully applied to the research of ultrafast fiber lasers owing to its unique 2D structure and excellent saturation absorption characteristics [9]. Inspired by this reason, researchers began to explore the new 2D structural materials with easy preparation, adjustable band gap and excellent nonlinear optical properties (i.e., topological insulators, Mxene, and transition metal dichalcogenides (TMDCs)) [10-13].

Recently, TMDC materials such as MoS₂, MoSe₂ and WSe₂ have been reported to have excellent saturable absorption properties. VSe₂ as one of the TMDC materials, it not only has similar characteristics as other TMDC materials including excellent nonlinear optical response and effective optical modulation, but also has more special characteristics. Each VSe₂ molecule consists of a metal V atom and two Se atoms, where the metal V atom is sandwiched between two Se atoms. Its layered structure of VSe₂ is similar to graphene, which also has the advantages of large specific area, metallic properties, high electrical conductivity, and excellent photoelectric properties [14, 15]. Moreover, the change of interlayer coupling will lead to the change of band structure and exhibit stronger optical absorption [16,17]. In addition, VSe₂ has broadband absorption spectrum characteristics due to its zero-band gap, thus it has excellent nonlinear optical response, strong light-material interaction, and ultrafast carrier dynamics. Therefore, VSe₂ possesses promising applications in optoelectronic devices, especially the nonlinear properties of this material have received much attention [18]. For example, Li et al. used VSe₂ as SA for the first time to achieve an ultrashort pulse with a pulse width of 910 fs [19]. Wang et al. realized a passively Q-switched laser based on VSe₂ [20]. H. Ahmad et al. realized 1.9 μm mode-locking operation based on VSe₂ SA material. Although, there have been reports in the literatures based on VSe₂ mode locking, the pulse output characteristics of VSe₂-based fiber lasers need to be improved. According to reports, the use of nanomaterial composites for mold locking can achieve superior performance than a single nanomaterial, for example, it has been reported in the literature that $MoS_2$/graphene nanocomposite enhances nonlinear optical response [21], and graphene/$WS_2$ nanocomposite realizes single-wavelength and dual-wavelength switchable mode locking [22]. Therefore, in order to obtain superior nonlinear optical performance than pure $VSe_2$ material, $VSe_2$ and graphene oxide (GO) are prepared into nanocomposites for mode-locked lasers. The reason for choosing GO is that GO nanomaterials have similar photoelectric properties to grapheme, and have the advantages of higher specific surface area, easier modification, and simpler preparation process. Furthermore, by anchoring $VSe_2$ on GO nano film, the excellent characteristics of $VSe_2$ and GO can be used to improve the mode locking performance. This has also prompted more new SA materials with broadband saturable absorption and high damage threshold to be explored and studied in erbium-doped fiber laser (EDFL). In addition to exploring individual 2D nanomaterials, heterostructures synthesized by stacking at least two materials have also garnered growing interest due to their unique physical, chemical, structural characteristics, and potential applications in optoelectronic devices. Compared with individual 2D devices, the van der Waals heterostructure shows significant advantages in terms of function and performance, which also provides a new idea for the research of new 2D materials in ultrafast fiber lasers [23,24].

On the other hand, the soliton molecules are special soliton states composed of several solitons with a specific pulse modulation interval. Compared with conventional soliton, soliton molecules can provide a new coding scheme to improve the communication capacity. Thus, it has potential applications in the fields of all-optical communication, pulse train processing, and time-resolved detection [25-27]. In recent years, many studies have been carried out on soliton molecules. For example, Peng et al. have studied the generation of soliton molecules in a dissipative system by using dispersive Fourier detection technology [28]. Hu et al. proposed a method of orbital angular momentum (OAM) analysis to map the internal phase evolution of soliton molecules into optical vortex motion and realize the visualization of complex phase dynamics in the structure of soliton molecules [29]. Numerical simulation, as a research method that can supplement the experiment, is often used to explore some dynamic evolution processes of solitons that are difficult to measure experimentally. For example, in recent years, the soliton dynamics in mode-locked lasers has attracted great attention. Among them, the use of numerical simulation to reveal the dynamics of multi-soliton mode-locking has also been extensively studied [30]. Including the beating and relaxation oscillations in the accumulation process from noise to steady-state soliton, as well as the phase shift during the interaction or dissociation of multiple soliton or soliton molecules [31, 32]. Through numerical simulation, it is possible to intuitively explore the complex soliton dynamics that are currently not easy to observe. Therefore, numerical simulation is of great significance for understanding basic soliton dynamics and promoting the development of experiments.

In this article, the $VSe_2$/GO nanocomposite is used as a SA device to achieve ultrashort pulse, and the dynamic behavior of soliton generation is studied. $VSe_2$/GO SA has excellent nonlinear characteristics, demonstrating a modulation depth of 14.3% and a saturated absorption intensity of 0.93 $MW/cm^2$. Integrating $VSe_2$/GO SA into the EDFL realizes a stable conventional soliton mode-locking operation at 1557.9 nm. It has an ultrashort pulse with a pulse width of 573 fs, a signal-to-noise ratio (SNR) of 60.4 dB, and a slope efficiency of 16.16%. Moreover, by adjusting the polarization controller (PC) and pump power, the soliton molecules based on $VSe_2$/GO nanocomposite are found for the first time. The formation and dynamical evolution process of conventional solitons and soliton molecules are further revealed by the numerical simulation. The spectral characteristics of solitons obtained by the simulation are in good agreement with the experimental results. These results indicate that the $VSe_2$/GO SA device has potential application value in the field of nonlinear optics.

# 2 Preparation and characterizations of SA material

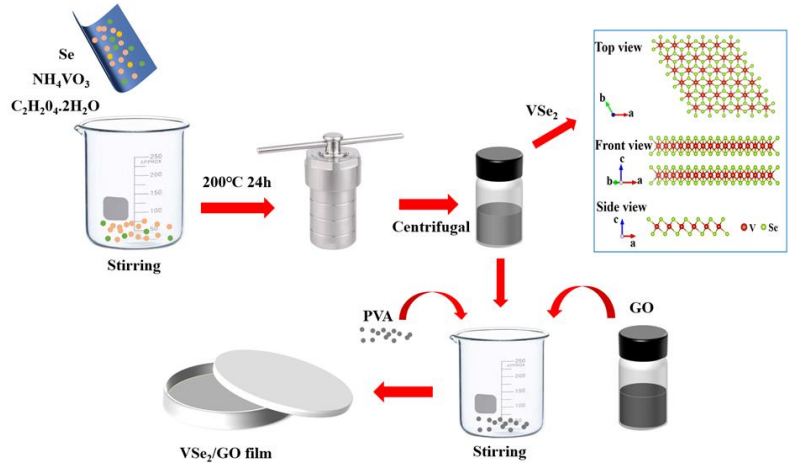

**Figure 1:** The preparation process of VSe₂/GO nanocomposite.

The preparation process of VSe₂/GO nanocomposite is shown in Figure 1. Firstly, 0.117 g of ammonium vanadate ($NH_4VO_3$, 1 mmol) and 1.198g of oxalic acid dihydrate ($C_2H_2O_4.2H_2O$, 9.5 mmol) were mixed in deionized water (40 ml) and stirred for 30 min. Then 0.158 g of Se powder (2 mmol) was added to the above solution and stirred continuously for 1 hour. Secondly, the above solution was reacted in an autoclave at 200°C for 24 hours. The VSe₂ precipitate was centrifuged several times with deionized water and anhydrous ethanol. The obtained VSe₂ product has a layered structure, and its molecular structure is illustrated in Figure 1. Thirdly, 5ml GO (0.2 mg/ml, modified Hummers' method) and 14ml VSe₂ (1 mg/ml) were mixed in deionized water (40 ml) and ultrasonic for 2 hours. Finally, 1g Polyvinyl alcohol (PVA) was added and stirred continuously for 1 hour, then took an appropriate amount of the solution and dried it in a polytetrafluoroethylene petri dish at 70°C for 6 hours.

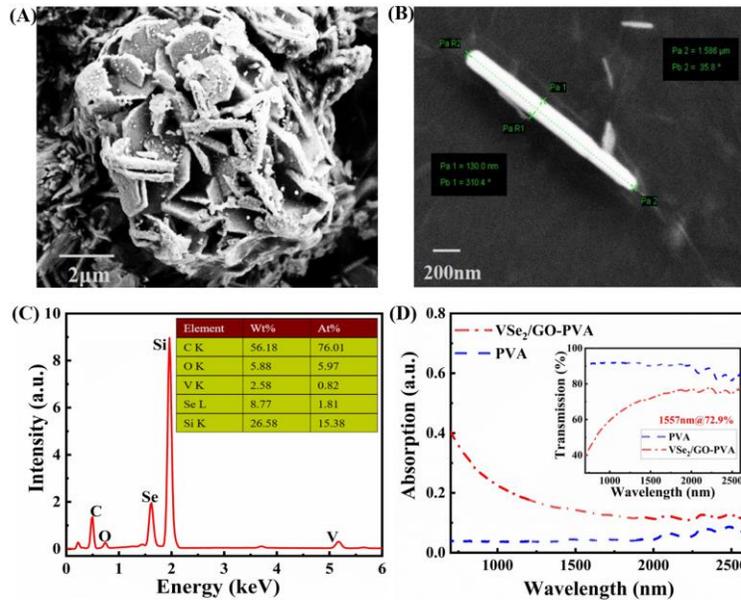

**Figure 2: Characterization and measurement of materials.** (A, B) SEM image of VSe₂/GO. (C) EDS spectrum of VSe₂/GO nanosheet. (D) Linear absorption spectrum of VSe2/GO SA. (Inset: the nonlinear optical absorption properties of VSe₂/GO).

Figures. 2(A) and 2(B) show the typical morphology of VSe₂/GO nanocomposite. The prepared VSe₂ presents a layered structure with a thickness between 100 nm and 200 nm. The thickness of a typical individual VSe₂ nanosheet shown in Figure 2(B) is 130 nm. Due to the package of GO, some flakes of VSe₂ gather. The energy dispersive spectroscopy (EDS) spectrum shown in Figure 2(C) confirms the existence of the elements Se, V, C, and O. In addition, the linear absorption and linear optical transmission of VSe₂/GO are also studied, as shown in Figure 2(D). It indicates that the VSe₂/GO thin film has broad spectral absorption so that it can be used as an SA device in EDFL. The absorption spectrum of PVA is flat, and the absorption value is small, which has little influence on the optical properties measurement and can be ignored. The illustration shows that the transmittance of VSe$_2$/GO thin film at 1557 nm is 72.9%, which is equivalent to that of other materials.

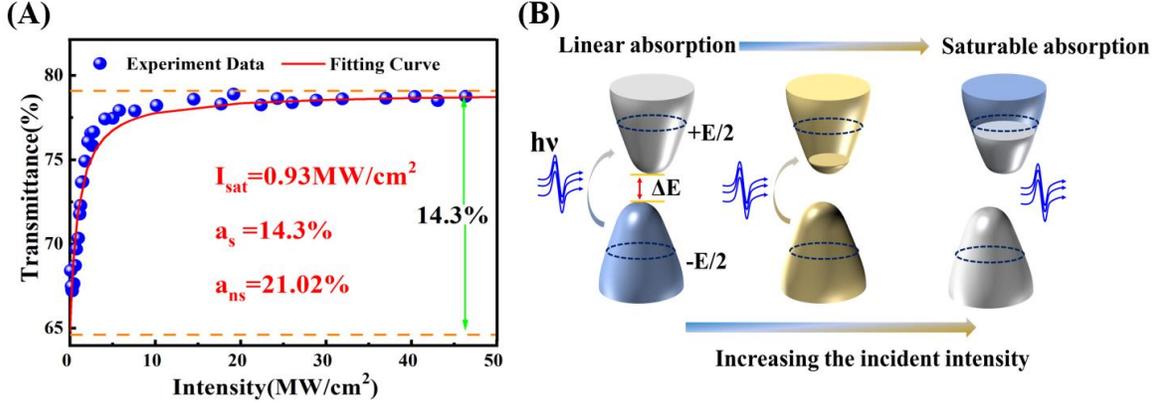

**Figure 3: Saturable absorption properties of SA.** (A) Nonlinear saturable absorption curve. (B) Three stages of the SA mechanism.

The double-balance detection system measured the saturated absorption curve of VSe$_2$/GO SA. The SA characteristics of nonlinear optical materials show that as the incident light intensity increases, the light absorption decreases. The following formula can express the saturated absorption function,

$$\alpha(I) = \frac{\alpha_s}{1 + I / I_{sat}} + \alpha_{ns},$$

where $\alpha(I)$, $\alpha_s$, $I_{sat}$ and $\alpha_{ns}$ denote the absorption, modulation depth, saturated absorption intensity, and nonsaturable loss, respectively. The nonlinear transmittance curve based on VSe$_2$ SA can be fitted with the $1-\alpha(I)$ function, as shown in Figure 3(A). The modulation depth and saturated absorption intensity of the SA are 14.3% and 0.93 MW/cm$^2$, respectively. In addition, the schematic energy band diagram is illustrated in Figure 3(B) to explain the underlying mechanism. When incident light is applied at high intensity, electrons in the valence band absorb the energy of photon and transition to the conduction band of SA material. Electrons and holes entirely occupy the band. According to Pauli's principle of incompatibility, when the absorption of VSe$_2$/GO material reaches saturation, photons can pass through without damage. That is, when the pulse passes through SA, weak light will be absorbed, and high-intensity light will cause the VSe$_2$/GO material to be bleached [33]. It is reported that the work function of VSe$_2$ is about 5.52eV-5.76eV [34-36], while the Fermi level of GO is about 4.9eV-5.5eV [37-39]. According to the different work functions of VSe$_2$ and GO, light-induced electrons are transferred from GO to VSe$_2$. The metallic properties of VSe$_2$ and the GO framework ensure the high electron transfer rate between composite materials [40]. This structure provides a channel for light-excited carrier transfer and enhances the interaction between light and electrons, which improve the nonlinear response properties. In addition, the enhanced optical absorption strength could benefit to stronger optical nonlinearity and further lead to larger modulation depth of VSe$_2$/GO. The larger modulation depth indicates that the VSe$_2$/GO SA has stronger bleaching ability to strong light, so a narrower mode-locked pulse width can be obtained. Therefore, the VSe$_2$/GO composite is suitable for SA devices to generate mode-locked pulse with excellent performance.

# 3 Ultrafast laser applications

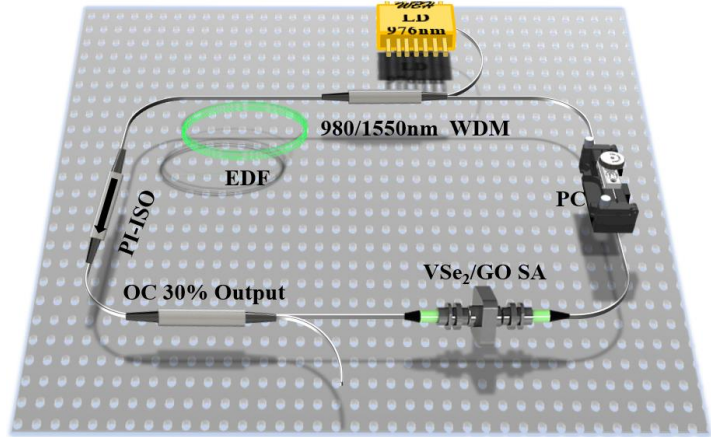

**Figure 4:** Diagram of the resonator device of passively mode-locked EDFL.

The schematic diagram of the all-fiber EDFL is shown in Figure 4. The laser cavity comprises a 6.3 m single-mode fiber and 0.4 m erbium-doped gain fiber (LIEKKI, Er110-4/125). The resonant cavity uses a 976 nm laser diode (LD) as the pump light source. The PC and the polarization-independent isolator (PI-ISO) in the resonant cavity are used to control the polarization state and ensure the one-way transmission, respectively. The VSe$_2$/GO SA is coupled into the resonant cavity in a "sandwich" between two fiber ferrules. Finally, the signal source to be detected is output through a OC, as shown in Figure 3. Its performance can be measured by an oscilloscope spectrum analyzer (Yokogawa AQ6370B), a radio frequency (RF) spectrum analyzer (Agilent N9020A), and a commercial autocorrelator (APE Pulsecheck).

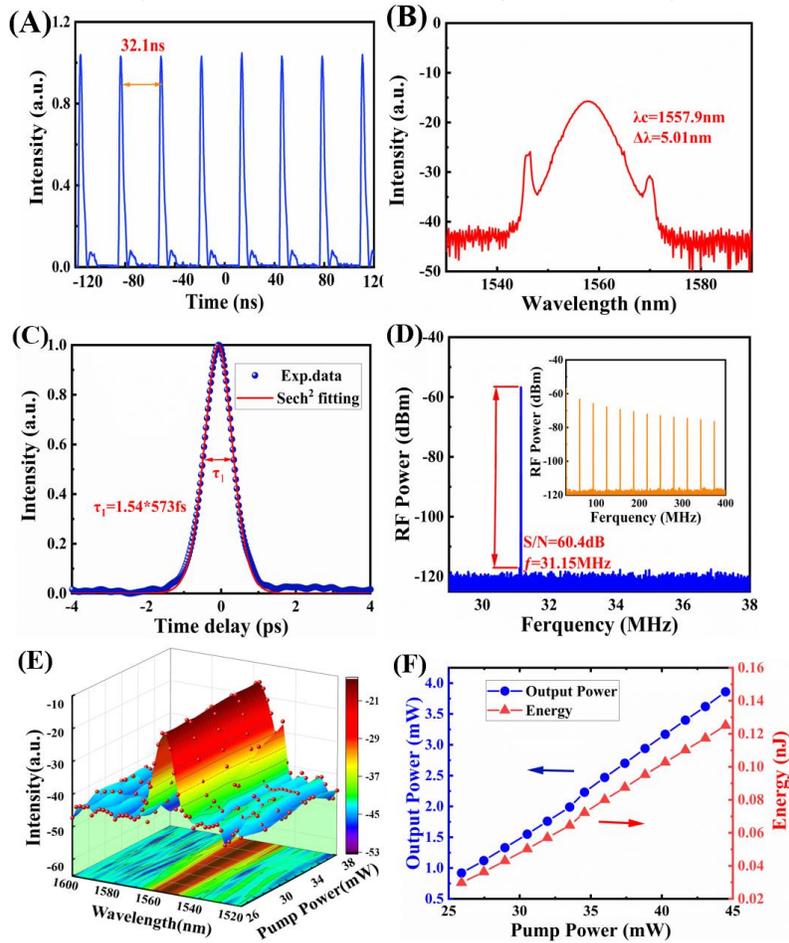

**Figure 5: Output performance of mode-locked pulse.** (A) Oscilloscope trace. (B) Optical spectrum. (C) Autocorrelation trace. (D) RF spectrum and the broadband RF spectrum (inset). (E) Spectral variation with pump power. (F) Output energy as a function of pump power.

The output characteristics of the mode-locked pulse are shown in Figure 5. When the pump power is increased to 26 mW, the mode-locked ultrashort pulse is obtained by adjusting the polarization state in the optical path. When the power is increased to 38 mW, the oscilloscope trace of the pulse sequence is shown in Figure 5(A). The interval between adjacent pulses is 32.1ns, and the fundamental repetition frequency is 31.15 MHz. The center wavelength of the pulse shown in Figure 5(B) is located at 1557.9 nm, and the 3 dB spectral bandwidth is 5.01 nm. The obvious Kelly sideband proves that the ultrashort pulse is a conventional mode-locked pulse. Figure 5(C) depicts the autocorrelation curve of the obtained mode-locked pulse, and fitting it with the sech$^2$ curve shows that the actual pulse width is ~573 fs. The time-bandwidth product (TBP) calculated by the following formula $TBP = c \cdot \frac{\Delta t \cdot \Delta \lambda}{\lambda^2}$ ( $c$ is the speed of light in vacuum, $\Delta t$ is the pulse width, $\Delta \lambda$ is the 3 dB bandwidth of the mode-locked laser spectrum, $\lambda$ is the central wavelength of the laser) is 0.355, indicating that there is a slight chirp phenomenon. As shown in the RF spectrum of Figure 5(D), it has a SNR of ~60.4 dB at a repetition frequency of 30.84 MHz, indicating that the mode-locking state is relatively stable. From Figure 5(E), it can be observed that the Kelly sideband of the pulse spectrum is gradually apparent with the increase of power. This is because the pulse energy spillovers when the power is increased, which is also the characteristic of a conventional mode-locked pulse. It can be seen from Figure 5(F) that the output energy of the resonant cavity increases linearly with the increase of the pump power. The slope efficiency of the mode-locked pulse is calculated to be 16.16%, which again indicates that VSe$_2$/GO is an excellent SA device. When the pump power changes, the nonlinear balance of the cavity will be affected, leading to the split of the mode-locked pulse and the appearance of modulation phenomenon among multiple pulses. The cross-phase modulation between pulses can result in the formation of soliton pairs. When the modulation reaches equilibrium, the phase relationship between multiple solitons is fixed, so the phenomenon of soliton molecules appears [25, 41-43]. In our work, soliton molecules are obtained by carefully adjusting the pump power and PC, as shown in Figure 6. The autocorrelation test shows that the pulse interval of the soliton molecules phenomenon is ~2.2 ps, which is basically consistent with the spectral modulation period of ~3.4 nm. And the soliton molecules have a SNR of ~53 dB measured by a RF spectrum, demonstrating excellent stability. The slight change in the repetition frequency (30.84 MHz) is the result of the change in the cavity length during the welding process. When VSe$_2$/GO is removed, the above-mentioned mode-locked pulse cannot be generated, proving that the optical nonlinearity of VSe$_2$/GO SA is the critical reason for the mode-locked phenomenon. Therefore, VSe$_2$/GO is an ideal SA material for ultrashort pulse generation.

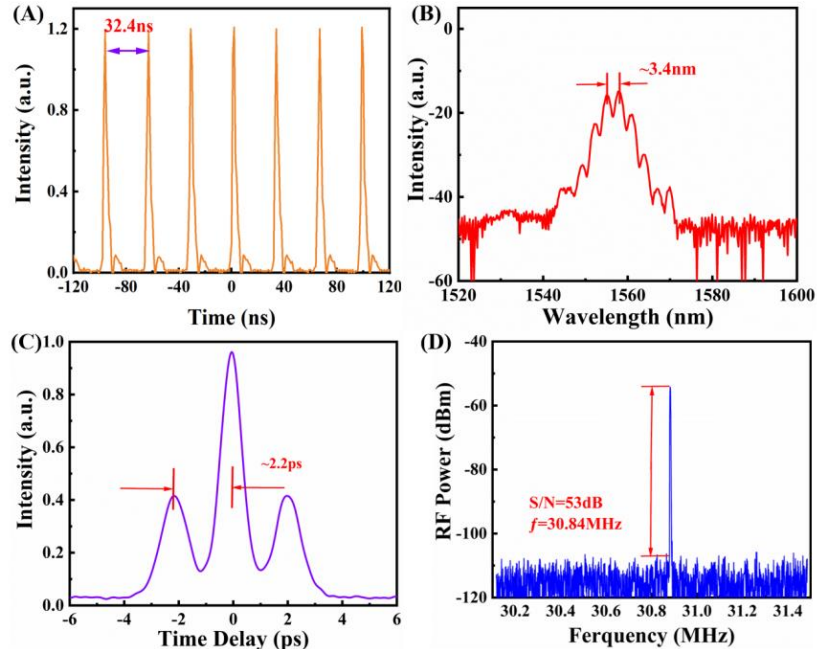

**Figure 6: Output performance of soliton molecules.** (A) Oscilloscope trace. (B) Optical spectrum. (C) Autocorrelation trace. (D) RF spectrum.

The performance output of some current mainstream 2D SA mode-locked devices in PML fiber lasers are listed in Table 1. As shown in the Table 1, compared with the current mainstream materials such as MoS2 and WS2, the pulse output based on VSe$_2$/ GO has advantages in pulse width and repetition rate. Moreover, compared with the reported results of pure VSe2 SA mode locking, our VSe$_2$/ GO-SA has a shorter pulse width of 573 fs and a larger repetition rate of 31.15 MHz in the mode-locked laser. In short, the comparison further proves that the VSe$_2$/GO SA device has excellent application potential in ultrafast laser.

**Table 1:** Output results of the fiber lasers based on different 2D material SAs

| Materials | Wavelength (nm) | Pulse Width (fs) | Frequency (MHz) | SNR (dB) | References. |
|---|---|---|---|---|---|
| MoS$_2$ | 1568.9 | $1.28 \times 10^3$ | 8.29 | 62 | [44] |
| WS$_2$ | 1558 | 830 | 8.2 | 70 | [45] |
| Ni-MOF | 1563.79 | 749 | 6.47 | 58 | [46] |
| VSe$_2$ | 1912 | $1.4 \times 10^3$ | 11.6 | 47 | [47] |
| VSe$_2$ | 1064.03 | $5.66 \times 10^6$ | 0.03 | 57.2 | [20] |
| VSe$_2$ | 1565.69 | 910 | 8.12 | 76 | [19] |
| Mo$_2$C/Graphene | 1599 | 723 | 15.33 | 68.6 | [48] |
| MoS$_2$/Graphene | 1596.2 | $1.36 \times 10^3$ | 9.8 | 73 | [24] |
| WS$_2$/Graphene | 1601.9 | 660 | 21.78 | 68 | [49] |
| Bi$_2$Te$_3$/FeTe$_2$ | 1558.8 | 481 | 23 | 55 | [10] |
| VSe$_2$/GO | 1557.5 | 573 | 31.15 | 60.4 | This work |

# 4 Simulation results and discussion

In order to study the dynamic evolution and the compression process of the intracavity pulse in the above-mentioned VSe$_2$/GO SA-based mode-locked laser, the numerical simulation is carried out based on the split-step Fourier method. The transmission of pulses in optical fiber systems can be modeled theoretically using generalized nonlinear Schrödinger equation given as follows [28,50,51],

$$\frac{\partial \psi(z,t)}{\partial z} = \frac{(g-\alpha)}{2}\psi(z,t) + \left(\frac{g}{2\Omega_g^2} - i\frac{\beta_2}{2}\right)\frac{\partial^2 \psi(z,t)}{\partial t^2} + i\gamma |\psi(z,t)|^2 \psi(z,t), \quad (2)$$

where $\psi(z,t)$ is the pulse envelope, $\beta_2$, $\gamma$ and $\Omega_g$ is the second-order dispersion, Kerr nonlinearity coefficient and bandwidth of the gain, respectively. $\alpha$ is the loss coefficient and g is the gain coefficient given as follows,

$$g = g_0 \exp\left(-\frac{E_{pulse}}{E_{sat}}\right), \quad (3)$$

where $g_0$, $E_{pulse}$ and $E_{sat}$ represent small-signal gain, pulse energy and saturation energy, respectively. In the numerical simulation, the initial input pulse is a weak Gaussian pulse and the fiber parameters used in the experiment are adopted, which are given as follows : EDF=0.4m, SMF=6.3m, $\alpha = 0.2dB \cdot km^{-1}$, $g_0 = 8m^{-1}$, $\Omega_g = 30nm$, $\beta_2 \approx 13\,ps^2 \cdot km^{-1}$, $\gamma \approx 3.6 \times 10^{-3}W^{-1} \cdot m^{-1}$ for EDF and $g = 0, \alpha = 0.2dB \cdot km^{-1}$, $\beta_2 \approx -19.8\,ps^2 \cdot km^{-1}$, $\gamma \approx 2 \times 10^{-3}W^{-1} \cdot m^{-1}$ for SMF. The initial pulse propagates through the SA in the cavity is modeled by instantaneous transfer function $T(I)$, the transmission of the SA is modeled by

$$T(I) = 1 - \frac{\alpha_s}{1 + \frac{I(t)}{I_{sat}}} - \alpha_{ns}, \quad (4)$$

where $I(t)$ is the instantaneous pulse energy, $\alpha_s$, $I_{sat}$ and $\alpha_{ns}$ stands for the modulation depth, saturation absorption intensity, and nonsaturable loss, respectively.

Figure 7 shows the evolutionary process of the conventional soliton, when the saturation energy given is taken as $E_{sat}$ =15.5pJ .

Figure 7(A) shows the spectral profile when the round-trip (RT) is 1500 in the mode-locking, and its center wavelength is around 1560nm. It can be seen from the spatio-spectral dynamics in Figure 7(B) that the simulation spectrum changes from a narrow-band continuous wave to a mode-locked state with a broadband spectrum. Moreover, during the transition process of mode-locking, the characteristic of spectral dynamics is that the spectral intensity oscillates before the stable mode-locking.When the spectral intensity oscillation duration is about 150 RT, it is close to stable mode-locking. Corres pondingly, Figure 7(C) shows the pulse profile when RT=1500, and the full width at half maxima (FWHM) is approximately 879 fs, consistent with the experimental results (FWHM is 882fs). Figure 7(D) depicts spatio-temporal dynamics, for the initial mode-locked state, the pulse intensity continues to increase to a certain extent and then shows a decreasing trend, and finally evolves into a stable mode-locked pulse sequence. From the pulse energy evolution in Figure 7(E), before evolving into stable mode-locking, the pulse energy first increases and then undergoes a process of decreasing intensity, which is consistent with the spatio-temporal dynamics in Figure 7(D). In addition, the mode-locking process is composed of an energy accumulation stage and an unstable pulse stage [52, 53].

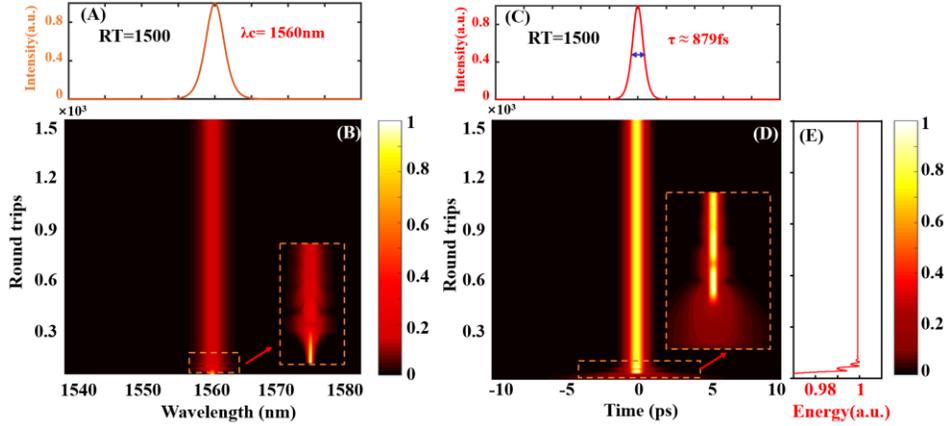

**Figure 7: Numerical simulation results of conventional soliton dynamics in 1500 consecutive round trips.** (A) Spectral profile with RT=1500. (B) Spatio-spectral dynamics. Inset: local magnification of the spectral domain. (C) Pulse profile with RT=1500. (D) Spatio-temporal dynamics. Inset: local magnification of initial pulse evolution. (E) Pulse energy evolution diagram.

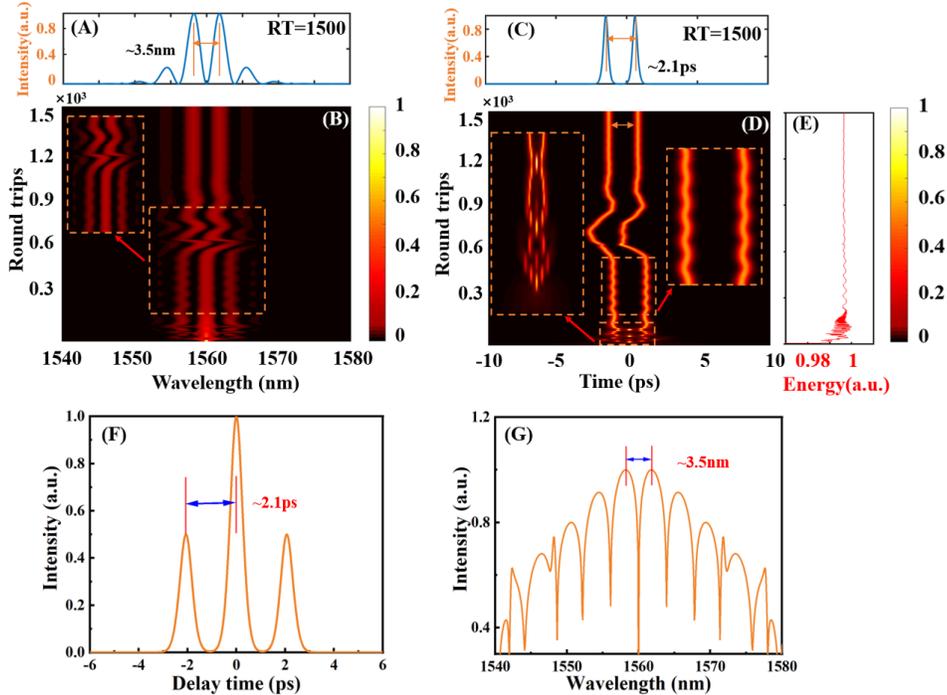

**Figure 8: Numerical simulation results of soliton molecules dynamics in 1500 consecutive RTs.** (A) Spectral profile with RT=1500, (B) Spatio-spectral dynamics. Inset: local magnification of the spectral domain. (C) Pulse profile with RT=1500. (D) Spatio-temporal dynamics. Inset: local magnification in the soliton interactions and vibration. (E) Pulse energy evolution diagram. (F) Pulse autocorrelation trace diagram with RT=1500. (G) Spectral profile with RT=1500 (log scale).

It is considering that $E_{sat}$ is the saturation energy determined by pump power in the experiment. Since the above simulation, an interesting soliton molecules phenomenon is obtained when $E_{sat}$=73.6pJ . The pulse evolution characteristics are shown in Figure 8. Figure 8(A) shows the spectral profile when RT=1500, and the result shows that the soliton molecules have a spectral modulation interval of about 3.5 nm. It can be seen from the spatio-spectral dynamics in Figure 8(B) that the evolves from a single pulse spectrum state to a mode-locked state with an apparent spectrum modulation and exhibits periodic spectrum interference phenomena within about 200-800 RTs. The interference image becomes obvious at RT=600, continues until RT=1000, and the spectrum shows steady propagation. Correspondingly, Figure 8(C) depicts the pulse profile when RT=1500. The separation interval between the two pulses is ~ 2.1 ps, which is consistent with the experimental result (~2.2 ps). Figure 8(D) depicts space-time dynamics. The pulse first undergoes a splitting process, it can be seen from the illustration on the left of Figure 8(D) that a single pulse is gradually split into multiple sub-pulses, and experienced multiple splits and synthesis [27,54,55]. In addition, modulation instability is an essential process of nonlinear science. It can also cause pulse splitting, resulting in multiple pulses [47]. And from the pulse energy evolution in Figure 8(E), before evolving into stable mode-locking, the pulse energy first increases to a certain degree and then exhibits periodic oscillation. The energy change is similar to the single pulse mode-locking. It gradually stabilizes from a high energy point. The changing trend of this energy is matched with the evolution of space-time dynamics. Figure 8(F) shows the autocorrelation trajectory of stable soliton molecules when RT=1500. It can be observed that the pulse separation interval (~2.1 ps) is in good agreement with the experimental result (~2.2 ps) of Figure 6(B). The modulation period (~3.5 nm) of the spectrogram (log scale) in Figure 8(G) matches the experimental result (~3.4 nm) of Figure 6(A) well. The generation process of soliton molecules is similar to previous literature reports [56].

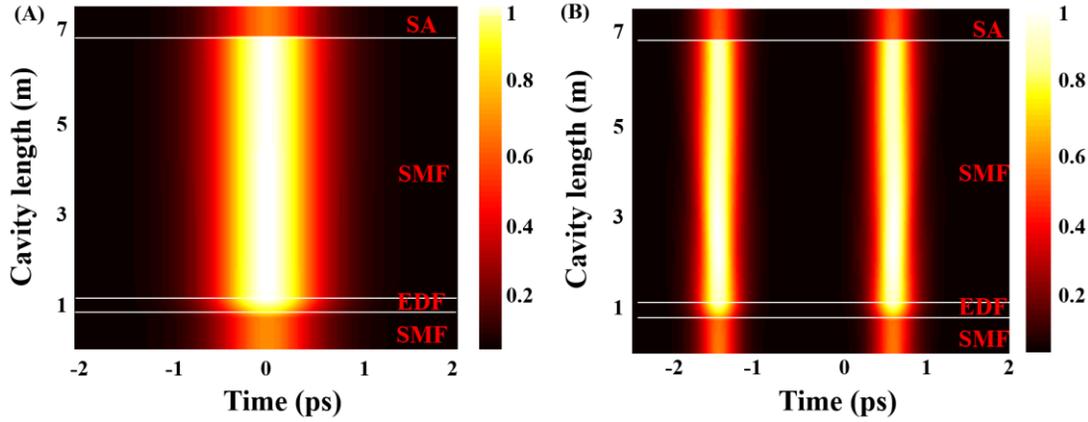

**Figure 9: Simulation results of the stable pulse evolution along the cavity**. (A) Conventional soliton. (B) Soliton molecules.

In order to further explore the formation mechanism of mode-locked pulse, we also qualitatively valuate the evolution of temporal along the cavity length, as shown in Figure 9. It is found that the pulse width and intensity of conventional solitons and soliton molecules that are in the evolution of temporal along the cavity length have similar variations. When the pulse enters the EDF from SMF, the pulse intensity show an increasing trend under the influence of gain, and then it is transmitted steadily in SMF. Finally, a saltation phenomenon is found at the position of the SA. Based on evolutionary dynamics, the saturation absorption and filter effect of VSe₂/GO SA are balanced with the gain term in the cavity, which is essential for the formation of stable solitons [57].

The reason for the slight difference between the above simulation results and the experiment may be caused by external interference in the experiment and high-order terms not considered in the numerical simulation model. Nevertheless, the accuracy is still outstanding. It shows the many nonlinear stages in the mode-locking process in fiber lasers and the generation of soliton molecules and other interesting nonlinear phenomena.

# 5 Conclusion

The VSe₂/GO nanocomposite is used as a SA device to achieve ultrashort pulse, and the dynamical behavior of soliton generation is studied. The VSe₂/GO SA shows excellent saturable absorption characteristics with a modulation depth of 14.3% and a saturation absorption intensity of 0.93 MW/cm². An ultrashort pulse with a pulse width of ~573 fs, a SNR of ~60.4 dB and a slope efficiency of 16.16% is achieved in the conventional soliton mode-locking region, showing a relatively high-quality pulse. The soliton molecules based on $VSe_2$/GO nanocomposite are found for the first time. Two identical solitons construct soliton molecules with a time interval of ~2.2 ps and a SNR of ~53 dB. The formation and dynamic evolution processes of conventional soliton and soliton molecules are further revealed by the numerical simulation, the result of the numerical simulation is consistent with experimental results. By studying the dynamics of these solitons will deepen the understanding of soliton pulsation in ultrafast fiber lasers. The above results show that the $VSe_2$/GO SA possesses excellent nonlinear characteristic and great potential as a SA in ultrafast fiber lasers.

**Acknowledgments**: B. H. Wang and H. B. Han contributed equally to this work. The research was supported by the Zhejiang Provincial Natural Science Foundation of China (Grant No. LR20A050001); National Natural Science Foundation of China (Grant Nos. 11874324,12075210); Scientific Research and Developed Fund of Zhejiang A&F University (Grant No. 2021FR0009).